\documentclass[11pt,a4paper]{article}

\usepackage{jheppub}
\usepackage{color}

\usepackage{graphicx}

\usepackage{ulem}

\newcommand{\nn}{\nonumber}
\def\dfrac#1#2{\displaystyle\frac{#1}{#2}}

\newcommand{\pslash}{p\kern-1ex /}
\newcommand{\qslash}{q\kern-1ex /}
\newcommand{\lslash}{l\kern-1ex /}
\newcommand{\sslash}{s\kern-1ex /}
\newcommand{\kaslash}{k_a\kern-2ex /}
\newcommand{\kbslash}{k_b\kern-2ex /}
\newcommand{\Dslash}{{\cal D}\kern-1.5ex /}

\newcommand{\tr}{{\rm tr}}

\newcommand{\beqa}{\begin{eqnarray}}
\newcommand{\eeqa}{\end{eqnarray}}

\begin{document}
\begin{flushright}
\normalsize{
YITP-22-03
\\OU-HET-1128}
\end{flushright}
\title{Conserved non-Noether charge in general relativity:
Physical definition vs. Noether's 2nd theorem}

\author[a]{Sinya AOKI,}
\affiliation[a]{Center for Gravitational Physics, Yukawa Institute for Theoretical Physics, Kyoto University, Kitashirakawa Oiwakecho, Sakyo-ku, Kyoto 606-8502, Japan}

\author[b]{Tetsuya ONOGI\,}
\affiliation[b]{Department of Physics, Osaka University, Toyonaka, Osaka 560-0043, Japan}

\abstract{In this paper, we make a close comparison of  a covariant definition of an energy/entropy in general relativity, recently proposed by a collaboration including the present authors,  with existing definitions of energies  such as the one from the pseudo-tensor and the quasi-local energy. 
We show that existing definitions of energies in general relativity are conserved charges from the Noether's 2nd theorem 
for the general coordinate transformation, whose conservations are merely identities implied by the local symmetry and always hold without using equations of motion.  Thus  none of existing definitions in general relativity reflects the dynamical properties of the system, need for a physical definition of an energy. 
In contrast, our new definition of the energy/entropy in general relativity is generically a conserved non-Noether  charge and  gives physically sensible results for various cases such as the black hole mass, the gravitational collapse, and the expanding universe, while existing definitions sometimes lead to
unphysical ones including zero and infinity.
We conclude that our proposal is more physical than existing definitions of energies. Our proposal makes it possible to  define almost uniquely the covariant and  conserved energy/entropy in general relativity, which brings some implications to future investigations. 
 }
  
\maketitle

 
 \section{Introduction}
 \label{sec:intro}
 Since Einstein proposed general relativity as a theory for gravity\cite{Einstein:1916}, 
 a proper definition of an energy, more generally a conserved charge from an energy momentum tensor~(EMT), 
 has been looked for. A main obstruction comes from a fact that a covariant conservation law 
 with a covariant derivative $\nabla_a$
 for an energy momentum tensor of matters $T^a{}_b$  in general relativity,
 \beqa
 \nabla_a T^a{}_b = 0,
 \label{eq:cov_convT}
 \eeqa
 is different from the standard  conservation law,
 \beqa
 \partial_a ( \sqrt{-g}  T^a{}_b ) =0,\quad g:=\det g_{ab},
 \label{eq:psT}
 \eeqa 
 which is required to construct a conserved energy but is not covariant under the general coordinate transformation, the most fundamental symmetry of general relativity.
 Einstein himself modified a definition of the energy momentum tensor as $\tilde T^a{}_b=T^a{}_b + t^a{}_b$
 to satisfy \eqref{eq:psT}.  
 Since $t^a{}_b$ is not a tensor under the general coordinate transformation except the affine transformation,
 $\tilde T^a{}_b$ is called Einstein's energy momentum {\it pseudo-tensor}.
 A more modern way is to define a total energy of a system by a surface integral of gravitational fields
 in its asymptotic region, called a quasi-local energy, for an asymptotically flat spacetime\cite{Arnowitt:1962hi,Bondi:1962px,Brown:1992br}.
 This approach has been extended further for more general  asymptotic behaviors by properly incorporating extra surface 
 terms\cite{Hawking:1995fd,Horowitz:1998ha,Balasubramanian:1999re,Ashtekar:1999jx}.  
 See \cite{DeHaro:2021gdv} for a recent summary of the problem including historical perspectives.
 
Recently, the present authors  and their collaborator have proposed a different definition for conserved charges such as the energy and its generalization in a curve spacetime including general relativity\cite{Aoki:2020prb,Aoki:2020nzm}, directly from the energy momentum tensor of matters but still keeping its covariance under the general coordinate transformation. 
Advantages of this definition, however, have not been fully recognized,
partly because our previous papers focused on the idea and the quick report of the results without detailed comparisons
to existing definitions.
Thus, in this paper, we make detailed comparisons between our proposal and other definitions for conserved charges in general relativity, showing that our definition is much more natural and  physical than others,
in order to establish that our definition of the energy and its generalization solves the long standing issue for the definition of the energy in general relativity. 

In Sec.~\ref{sec:2nd}, 
we demonstrate that (almost) all existing definitions of the energy in general relativity can be 
regarded as a conserved charge implied  by the Noether's 2nd theorem for local symmetries\cite{Noether:1918zz}. 
We show that definitions of the energy as charges from the Noether's 2nd theorem are categorized either 
as the Einstein's pseudo-tensor type or as the Komar energy\cite{Komar:1958wp} type, the later of which includes
the ADM mass\cite{Arnowitt:1962hi}, and the energy in the asymptotically flat spacetime\cite{Bondi:1962px,Brown:1992br} as well as in the asymptotically dS/AdS spacetime\cite{Hawking:1995fd,Horowitz:1998ha,Balasubramanian:1999re,Ashtekar:1999jx}. 
Since both types of definitions allow quasi-local  expressions, 
we can easily change their definitions of the energy by adding an arbitrary total divergent term to the Einstein-Hilbert action. 
Even worse, the energy from these two types of definitions is conserved {\it without} using equations of motion. 
Thus, the conservation of the energy  is merely identity implied by the general coordinate transformation rather than a consequence of a time 
evolution, so that it cannot represent a dynamics of the system.  
We conclude that none of existing definitions  from the Noether's 2nd theorem 
can provide a physical definition of an energy in general relativity.
Indeed Noether herself referred the charge from the 2nd theorem {\it improper} by citing the word from Hilbert and Klein\cite{Noether:1918zz}.  

In Sec.~\ref{sec:proposal}, we instead explain our proposal for a covariant definition of the energy and its generalization in general relativity, which requires equations of motion\cite{Aoki:2020prb,Aoki:2020nzm}, and thus is not a charge from the 2nd theorem.
After reviewing our proposal, we discuss three cases, (1) energy conservation by a global symmetry, (2) energy conservation without symmetry,  (3)   conserved charge in the absence of energy conservation, together with explicit examples, where we {also} compare results from our proposal with those from  the Noether's 2nd theorem.
In the case (1), our definition gives the finite energy of the Schwarzschild black hole even for non-zero cosmological constant $\Lambda$, while definitions from  the Noether's 2nd theorem require a subtraction of the infinite vacuum energy to obtain the finite black hole energy for $\Lambda\not= 0$ cases, which agrees with the one from our definition only at $d=4$.
We have a similar comparison for the energy during a gravitational collapse in the case (2). 
In the case (3), the homogeneous and isotropic expanding Universe is analyzed.
While the energy in our covariant definition is not conserved, we show that our definition allows a conserved charge as the generalization of the energy, which we identify the entropy.
On the other hand, the conservation of the energy for definitions from  the Noether's 2nd theorem  
implies the vanishing total energy, which is physically meaningless.

 Our conclusion and discussion are given in Sec.~\ref{sec:concl}. 
 For the sake of readers, the Noether's 2nd theorem is explained for general cases
in appendix~\ref{sec:Noether2nd}.

\section{Noether's 2nd theorem and conserved charges in general relativity}
\label{sec:2nd}
In this section, we derive conservation equations using Noether's 2nd theorem in general relativity.  
We then show that  these conservation equations lead to a pseudo-tensor as well as charges associated with asymptotic symmetry including the ADM mass. 
  
\subsection{Noether's 2nd theorem in general relativity}
We apply the Noether's 2nd theorem to  general relativity.
The Noether's 2nd theorem is given in \cite{Noether:1918zz}, and its application to general relativity is discussed
in   \cite{Utiyama:1984bc}, but these considerations, except the famous Noether's 1st theorem, have not been recognized well or 
have been sometimes misunderstood in the community.
Thus, for the sake of readers, we explain the 2nd theorem here in the case of general relativity, and the derivation of the theorem   is presented for a general case in the appendix~\ref{sec:Noether2nd}.

To make our argument concrete,  we take a scalar field theory coupled to the Einstein gravity, whose Lagrangian density is given by
\beqa
L &=& L_G + L_M
\eeqa
where
\beqa
L_G &=& \frac{1}{2\kappa} \sqrt{-g} (R - 2\Lambda), \quad \kappa:= 4\pi G,
\\
L_M &=& \sqrt{-g} \left[ -\frac{1}{2}g^{ab} \partial_a \phi \partial_b \phi - V(\phi) \right],
\eeqa 
and consider the integral of $L$ over an arbitrary $d$-dimensional region $\Omega$ in the $d$-dimensional spacetime as
\beqa
S_\Omega &:=& \int_\Omega d^d x\, L.
\eeqa  

We first derive an equation of motion by considering an arbitrary variation $\delta_v$ as
\beqa
2\kappa \delta_g S_\Omega &=&\int_\Omega d^d x\, \sqrt{-g} \left[ \left(\frac{1}{2} g^{ab} (R-2\Lambda) - R^{ab}+2\kappa  T^{ab} \right)
 \delta_v g_{ab} 
+ \nabla_a \left( g^{bc} \delta_v \Gamma^a_{bc} - g^{ab} \delta_v \Gamma^c_{bc} \right)
 \right], \nn \\
\delta_\phi S_\Omega &=& \int_\Omega d^d x\,  \left[ \sqrt{-g}\left(\nabla_a \nabla^a \phi -V^\prime(\phi) \right)\delta_v \phi -\partial_a \left(\sqrt{-g} g^{ab} \partial_b \phi   \delta_v \phi \right) \right] ,
\eeqa
where
\beqa
T^{ab} &:=& \frac{1}{\sqrt{-g}}\frac{\partial L_M}{\partial g_{ab}} ={1\over 2}\left[
\partial^a\phi \partial^b \phi -\frac{1}{2}g^{ab}\left(\partial^c\phi \partial_c\phi + 2 V(\phi) \right)\right],
\eeqa
and we use a fact that $\delta_v \Gamma^a_{bc}$ can be regarded as a mixed tensor.
Since we can take arbitrary variations which, together with  their derivatives, 
vanishes at the boundary of $\Omega$, we obtain equations of motion as
\beqa
E^{ab}_G&:=& -\frac{\sqrt{-g}}{2\kappa} \left(R^{ab}-\frac{1}{2} g^{ab} (R-2\Lambda) - 2\kappa  T^{ab}\right)=0, \\ 
E_\phi&:=&\sqrt{-g}\left(\nabla_a \nabla^a \phi -V^\prime(\phi)\right) = 0.
\eeqa 
Note that we can add the total derivative term $\partial_a (\sqrt{-g} K^a)$ to the Lagrangian density $L$ without changing equations of motion. Thus,  there is an ambiguity for a choice of the Lagrangian density from which we can derive above equations of motion. 
In our analysis we exclusively use the above $L$, keeping this ambiguity in mind.
In particular, we take the Einstein-Hilbert type for $L_G$.

We now consider a general coordinate transformation generated by $\xi^a$ as
\beqa
\delta x^a&:=& (x^\prime)^a - x^a =\xi^a (x), \quad \delta\phi :=\phi^\prime(x^\prime) - \phi(x)=0 , \nn \\
\delta g_{ab}& :=& g^\prime_{ab}(x^\prime) -g_{ab}(x)=- \xi^c{}_{,a}(x) g_{cb}(x) -   \xi^c{}_{,b}(x) g_{ac}(x).
\eeqa
Since $\delta$ does not commute with derivatives, we introduce the Lie derivative by $\xi$ as
\beqa
\bar\delta g_{ab} := \delta g_{ab} - g_{ab,c} \xi^c = -\nabla_a\xi_b -\nabla_b\xi_a,\quad
\bar\delta \phi := \delta\phi -\xi^c\phi_{,c} =-\xi^c \nabla_c \phi, 
\eeqa 
which satisfies
\beqa
\bar\delta( g_{ab, c\cdots}) = (\bar\delta g_{ab})_{,c\cdots}, \quad
\bar\delta( \phi_{, c\cdots}) = (\bar\delta \phi)_{,c\cdots}.
\eeqa
A fact that an integration of the Lagrangian density over a $d$-dimensional domain $\Omega$ 
is invariant under the general coordinate transformation 
leads to
\beqa
\delta S_\Omega &=& \int_{\Omega} d^dx\, \left[\delta( L_G+ L_M ) + (L_G+L_M) \xi^a_{,a} \right]\nn \\
&=&\int_{\Omega} d^dx\, \left[\bar\delta( L_G+ L_M ) + \partial_a\{(L_G+L_M) \xi^a\} \right] = 0,
\label{eq:delS}
\eeqa
where we employ
\beqa
d^d (x+\delta x) = \det \left[ \delta^a_b + (\delta x^a)_{,b}\right] d^d x 
\simeq (1 + \tr\, \xi^a_{,b} ) d^d x = (1+ \xi^a_{,a} ) d^d x , \\
\delta(L_G+L_M) = \bar\delta (L_G+L_M) + \xi^a\partial_a(L_G+L_M)  .
\eeqa

Using 
\beqa
\bar\delta (L_G+L_M) &=& \left(E^{ab}_G\bar\delta g_{ab} + E_\phi \bar\delta\phi \right) 
+\partial_a \left\{ \frac{\sqrt{-g}}{2\kappa}\left( g^{bc}\bar\delta \Gamma^a_{bc} - g^{ab}\bar\delta \Gamma^c_{bc}- 2\kappa g^{ab} \partial_b \phi   \bar\delta \phi \right)
\right\}, \nn \\
\eeqa
and 
\beqa
E_G^{ab}\bar\delta g_{ab}&=&\xi^c\left[ 2\partial_a\left( E^{ab}_G g_{bc}\right) - E_G^{ab} g_{ab,c}
\right] 
-2\partial_a \left( E^{ab}_G g_{bc}\xi^c \right), 
\eeqa
we have
\beqa
\delta S_\Omega &=&  \int_{\Omega} d^dx\,
\xi^c\left[ 2\partial_a\left( E^{ab}_G g_{bc}\right) - E_G^{ab} g_{ab,c} 
- E_\phi \nabla_c \phi \right]
+\int_\Omega d^dx\, \partial_a J^a[\xi ]=0,  
\label{eq:deltaS}
\eeqa
where
\beqa
J^a[\xi] &=& (L_G + L_M) \xi^a - 2E^{ab}_G g_{bc}\xi^c 
+  \frac{\sqrt{-g}}{2\kappa}\left( g^{bc}\bar\delta \Gamma^a_{bc} - g^{ab}\bar\delta \Gamma^c_{bc}- 2\kappa g^{ab} \partial_b \phi  \bar\delta \phi \right)\nn \\
&=& \frac{1}{2\kappa}\sqrt{-g}  \left[2 R^a{}_b\xi^b   
+g^{bc}\bar\delta \Gamma^a_{bc} - g^{ab}\bar\delta \Gamma^c_{bc}\right] 
=  \frac{1 }{2\kappa}\sqrt{-g}\nabla_b\left[ \nabla^{[a}\xi^{b]}\right].
\label{eq:Ja}
\eeqa
To obtain the last line, we use $R^{a}{}_b\xi^b = - g^{ac}[\nabla_c,\nabla_b]\xi^b$, 
\beqa
g^{bc}\bar\delta \Gamma^a_{bc} &=& -g^{bc}\nabla_b \nabla_c \xi^a + g^{ac}[\nabla_c, \nabla_b] \xi^b, \quad
g^{ab} \bar\delta \Gamma^c_{bc} =-g^{ab} \nabla_b\nabla_c\xi^c.
\eeqa

Since  we can take an arbitrary vector field $\xi_a(x)$ which satisfies
$\xi_a=\xi_{a,b}=\xi_{a,bc}=0$ at $\partial \Omega$ (the boundary of the region $\Omega$) as a general coordinate transformation, 
\eqref{eq:deltaS} implies 
\beqa
2\partial_a\left( E^{ab}_G g_{bc}\right) - E_G^{ab} g_{ab,c} 
- E_\phi \nabla_c \phi = 0,
\label{eq:constEOM}
\eeqa 
for {\it off-shell} $g_{ab}$ and $\phi$, which give $d$ constraints among 
the quantities
$E^{ab}_G$ and $E_\phi$, which would vanish at on-shell, so that solutions to the equation of motion contain $d$ undetermined free functions. 
In other words, \eqref{eq:constEOM} identically holds.
Thus a number of independent components for the symmetric tensor $g_{ab}$ becomes $d(d+1)/2 - d =d(d-1)/2$, as is well known.

Furthermore, taking an arbitrary $\xi_a(x)$ without constraints on $\partial \Omega$,
\eqref{eq:deltaS} with \eqref{eq:constEOM} leads to 
\beqa
 \partial_a J^a[\xi] &=& 0,  
 \label{eq:convJ}
\eeqa
where $J^a[\xi]$ includes the arbitrary vector $\xi^a$. 
Indeed, we can confirm that $\partial_a J^a[\xi]=0$ holds identically using an explicit form of $J^a[\xi]$
in the last line of \eqref{eq:Ja}. 

The current $J^a[\xi]$ is expanded as
\beqa
J^a[\xi] = A^a{}_b \xi^b + B^{a}{}_b{}^c \xi^b_{,c} + C^a{}_b{}^{cd}  \xi^b_{,cd}, 
\label{eq:conv_pt}
\eeqa
where
\beqa
A^a{}_b&=& {\sqrt{-g}\over 2\kappa}\left( 2R^a{}_b +g^{ca}\Gamma^{d}_{db,c} - g^{cd}\Gamma^{a}_{cd,b} \right) 
={\sqrt{-g}\over 2\kappa} \left[ \partial_c(g^{d[a}\Gamma^{c]}_{db}) + \Gamma^e_{ec} g^{d[a}\Gamma^{c]}_{db}\right],
\label{eq:Aab}\\ 
B^{a}{}_b{}^c &=&  {\sqrt{-g}\over 2\kappa} \left( g^{ac}\Gamma^d_{db}-2g^{dc}\Gamma^a_{db}+g^{de}\delta^a_b\Gamma^c_{de}\right), \\
C^a{}_b{}^{cd} &=&  {\sqrt{-g}\over 4\kappa} \left( g^{ac}\delta^d_b +g^{ad}\delta^c_b-2g^{cd}\delta^a_b\right)
= C^a{}_b{}^{dc},
\eeqa
and \eqref{eq:convJ} for an arbitrary $\xi^a$ implies 
\beqa
\partial_a A^a{}_b &=& 0, \label{eq:A}\\
A^a{}_b + \partial_c B^{c}{}_b{}^a &=& 0 ,\label{eq:AB}  \\
B^a{}_b{}^c + B^c{}_b{}^a +2\partial_d C^d{}_b{}^{ac} &=& 0,,\label{eq:BC} \\
C^a{}_b{}^{cd} + C^d{}_b{}^{ac} +C^c{}_b{}^{da} &=& 0 \label{eq:C}
\eeqa
Combining \eqref{eq:AB}, \eqref{eq:BC} and \eqref{eq:C}, we can generally write
\beqa
A^a{}_b &=& - {1\over 2} \partial_c B^{[c}{}_b{}^{a]} - {1\over 2} \partial_c B^{\{c}{}_b{}^{a\}}
=- {1\over 2} \partial_c B^{[c}{}_b{}^{a]} +\partial_c \partial_d C^{d}{}_b{}^{ac}= -\partial_c \tilde B^{c}{}_b{}^{a}, 
\label{eq:defA}
\eeqa
where
\beqa
\tilde B^{c}{}_b{}^{a} &:=&   {1\over 2} B^{[c}{}_b{}^{a]} -{1\over 3} \partial_d C^{[c}{}_b{}^{a]d},
\label{eq:tildeB}
\eeqa
which is anti-symmetric under $a\leftrightarrow c$. 

We fully utilize the fact that the general coordinate transformation is generated by an arbitrary vector field $\xi^a(x)$ to obtain
\eqref{eq:constEOM}, \eqref{eq:convJ}, \eqref{eq:A}--\eqref{eq:C}, which are the consequence of the Noether's 2nd theorem.

There are two remarks.
First of all, if we add the total derivative term $X:=\partial_a (\sqrt{-g} K^a)$ to $L$, 
its variation under $\delta$ becomes (See \eqref{eq:delS})
\beqa
\int_\Omega d^dx\, \left[ \bar\delta X + \partial_a ( X \xi^a) \right] 
= \int_\Omega d^dx\, \partial_a  \left[ \bar\delta (\sqrt{-g} K^a) +  X \xi^a \right] ,
\eeqa
which leads to a shift of $J^a[\xi]$ as
\beqa
J^a[\xi] &\to& J^a[\xi]+  \sqrt{-g} \left[ \xi^{[a} \nabla_b K^{b]} - K^{[a}\nabla_b \xi^{b]} \right],
\eeqa
where we use
\beqa
\bar\delta K^a &=& K^b \nabla_b \xi^a -\xi^b \nabla_b K^a, \quad \bar\delta \sqrt{-g} = -\sqrt{-g} \nabla_b \xi^b. 
\eeqa
Secondly, even though we can take  $\xi^a(x)= \xi^a_0$ with a constant vector $\xi^a_0$, 
we still have the Noether's 2nd theorem, so that the current associated with this symmetry is always conserved 
without using equations of motion.
 
Using \eqref{eq:convJ} and \eqref{eq:A}, we can define two types of conserved charges, one is covariant, the other is non-covariant, which will be explained below.
Their conservation, however, is an  {\it identity} implied by the general coordinate transformation, and holds without using equations of motion.

\subsection{Non-covaraint conserved charge from the Noether's 2nd theorem: pseudo-tensor}
The non-covariant {\it off-shell} conserved current density is given by
\beqa
A^a{}_b &:=&
 {\sqrt{-g}\over 2\kappa} \left( 2R^a{}_b +g^{ac}\Gamma^{d}_{db,c} - g^{cd}\Gamma^{a}_{cd,b}  \right), 
\eeqa 
and the conservation law $\partial_a A^a{}_b =0$ implies
\beqa
0&=& \int_M d^d x\, \partial_a A^a{}_b =\int_{\Sigma_2} (d^{d-1}x)_a \, A^a{}_b - \int_{\Sigma_1} (d^{d-1}x)_a \, A^a{}_b 
+ \int_{\partial M_s} (d^{d-1}x)_a \, A^a{}_b,
\label{eq:conv_law}
\eeqa
where $M$ is the $d$-dimensional spacetime whose boundary consists of $\partial M=\Sigma_1\oplus \partial M_s  \oplus  \Sigma_2$.
Here
$\Sigma_{1}$ and $\Sigma_2$ are past and future directed space-like surfaces, respectively, and
$\partial M_s$ is a time-like boundary of $M$.
If 
$\displaystyle 
\int_{\partial M_s} (d^{d-1}x)_a \, A^a{}_b =0
$,
we can define a conserved charge as
\beqa
Q_{{\rm pseudo},b} &=&  \int_{\Sigma} (d^{d-1}x)_a \, A^a{}_b,
\eeqa
since it does not depend on a choice of space-like surfaces $\Sigma_{1,2}$. 
We call  $Q_{{\rm pseudo},b}$ the non-covariant conserved charge, since 
$A^a{}_b$ is not covariant under the general coordinate transformation\footnote{
It is  only covariant under Affine transformation that $\xi^a(x):= m^a{}_b x^b - b^a$.}.
Furthermore, \eqref{eq:defA}  leads to a quasi-local expression of  $Q_{{\rm pseudo},b}$ as
\beqa
Q_{{\rm pseudo},b} &=&  -\int_{\partial\Sigma} (d^{d-2}x)_{ac} \, \tilde B^c{}_b{}^a,
\eeqa
where  the boundary of $\Sigma$ is denoted by a spatial surface $\partial\Sigma$. 
  
As already noted before, the conservation of $Q_{{\rm pseudo},b}$ is an identity, which is
not a consequence from the dynamics of general relativity,
since equations of motion are not required to show it. 
In addition, if the equation of motion for $g_{ab}$ ($E^{ab}_G =0$) is used,  $A^a{}_b$ becomes
\beqa
A^a{}_b &=& \sqrt{-g} ( T^a{}_b + t^a{}_b), \quad
t^a{}_b:=  {1\over 2\kappa}\left[ R^a{}_b +{R-2\Lambda\over 2} \delta^a_b + g^{ca}\Gamma^{d}_{db,c} - g^{cd}\Gamma^{a}_{cd,b}\right],
\label{eq:psTensor}
\eeqa
where $t^a{}_b$ is not covariant due to the last two terms.
In the case of the vanishing cosmological constant,
by adding an appropriate total divergent term $\partial_\mu (\sqrt{-g} K^a)$ to the total Lagrangian density,
$t^a{}_b$  can be transformed to the Einstein's gravitational pseudo tensor, 
which was claimed to represent the gravitational contribution.
A distinction between matter and gravitational field, however, seems ambiguous, since
$R^a{}_b$ and $R$ in $t^a{}_b$ are also expressed in terms of $T$ and $T^a{}_b$.

Using \eqref{eq:psTensor} for $b=0$, one may define the conserved energy as
\beqa
E_{\rm pseudo} &=& -\int_{\Sigma} [d^{d-1}x]_a\,  \sqrt{-g} ( T^a{}_0 + t^a{}_0)\left(= \int_{\partial\Sigma} [d^{d-2}x]_{ac} \, \tilde B^c{}_0{}^a\right),
\eeqa 
where a minus sign is introduced for $E_{\rm pseudo}$ to match the standard definition of the energy.
While Einstein interpreted the second contribution from his pseudo tensor $t^0{}_0$ as the energy of the gravitational field,
it depends on a choice of the coordinates due to its non-covariance, and it sometimes diverges.

\subsection{Covariant conserved charge from the Noether's 2nd theorem: Komar integral}
The second type of the conserved current is given by $J^a$ itself as 
\beqa
J^a[\xi] &=&  \frac{1 }{2\kappa}\sqrt{-g}\nabla_b\left[ \nabla^{[a}\xi^{b]}\right],
\eeqa
which satisfies $\partial_a J^a[\xi]=0$ for an arbitrary vector $\xi^b$.
Then one may define the covariantly conserved charge as
\beqa
Q_{\rm Komar}[\xi] := \int_\Sigma [d^{d-1}x]_a \,  J^a[\xi] &=&{1\over 2\kappa} \int_\Sigma [d^{d-1}x]_a \,  \sqrt{-g}\nabla_b\left[ \nabla^{[a}\xi^{b]}\right]\\
&=& {1\over 2\kappa} \int_{\partial\Sigma} [d^{d-2} x]_{ab} \,  \sqrt{-g} \nabla^{[a}\xi^{b]},
\eeqa 
where the second line is a quasi-local expression.
We call this charge the Komar integral, since the expression is identical to the one introduced by Komar\cite{Komar:1958wp}.
This charge is conserved not only for an arbitrary metric $g_{ab}$ but also for  an arbitrary vector $\xi^b$.
Thus, one may define various different charges depending on a choice of $\xi^b$.
We introduce several such charges used in literature.

\subsubsection{Komar energy}
If the spacetime allows a time-like Killing vector $\xi_K^a$, one may define the energy as a charge associated with the Killing vector  as
$E_{\rm Komar} = Q_{\rm Komar}[\xi_K]$, which we call Komar ``energy".
Explicitly
\beqa
E_{\rm Komar} &=& {1\over \kappa} \int_\Sigma [d^{d-1}x]_a \,  \sqrt{-g}R^a{}_b \xi_K^b\\
&=&  {1\over \kappa} \int_\Sigma [d^{d-1}x]_a \,  \sqrt{-g}\left[ 2\kappa\left(T^a{}_b \xi_K^b -{T\xi_K^a\over d-2}\right) +{2\Lambda\xi_K^a\over d-2}\right],
\eeqa
where we use the equations of motion to obtain the 2nd line, which 
shows that the Komar ``energy" does not lead to the standard definition of the energy in the limit of the flat spacetime.
A time-like Killing vector is given by  $\xi_K^a = -\delta^a_0$ for the stationary spacetime, for example, 
where the metric $g_{ab}$ does not depend on the time coordinate $x^0$.
Since $\xi_K^a = -\delta^a_0$ is constant, the Komar energy coincides with the energy from the pseudo tensor by definition: $E_{\rm Komar}=E_{\rm pseudo}$.
Note that the Komar ``energy" $E_{\rm Komar}$ is always conserved as a consequence of the Noether's 2nd theorem,
even though $\xi_K^a$ is not a Killing vector for a generic (non-stationary) spacetime. 

\subsubsection{Wald entropy}
It has been proposed to define the black hole entropy\cite{Wald:1993nt},
by choosing $\xi^a=t^a + \Omega_H \varphi^a$, where $t^a$ is the stationary Killing field, $\varphi^a$ is the axial Killing field, and $\Omega_H$  is the angular velocity of the horizon.
In Ref.~\cite{Wald:1993nt},  it is concluded that $\partial_a J^a[\xi]=0$ holds {\it when} the equations of motion are satisfied.
This statement is misleading, however, since a full power of the Noether's 2nd theorem was not employed to derive $\partial_a J^a[\xi]=0$ in Ref.~\cite{Wald:1993nt}.
As we have frequently mentioned, $\partial_a J^a[\xi]=0$ can be derived from the Noether's 2nd theorem for an arbitrary $\xi^b$ {\it without} using equations of motion or $g_{ab}$ and matters. 

\subsubsection{Asymptotically flat spacetime: ADM energy}
A asymptotically flat spacetime is defined as a spacetime whose metric satisfies the vacuum Einstein equation without cosmological constant at $x^2\to +\infty$ (large space-like separation). 
In this case, the conserved energy is defined in Cartesian coordinate as\cite{Arnowitt:1962hi}
\beqa
E_{\rm ADM} &:=& {1\over 4\kappa} \int_{+\infty} [d^{d-2}x]_{0 i} \,(\partial_j h_{ij}-\partial_i h_{jj}) , \quad h_{\mu\nu}:= g_{\mu\nu}-\eta_{\mu\nu},
\eeqa
which is called as the ADM energy (or mass), where $i,j$ run from 1 to $d-1$, $\eta_{\mu\nu}$ is the flat Minkowski metric, and
$\int_{+\infty}$  means that the integral is evaluated at $x^2\to +\infty$.

The ADM energy can be written in a covariant manner as\cite{Townsend:1997ku}
\beqa
E_{\rm ADM} &=& {1\over 4\kappa} \int_{+\infty} [d^{d-2 x}]_{ab} \, \sqrt{-g} \nabla^{[a}\eta^{b]} ={1\over 2} Q_{\rm Komar}[\eta],
\eeqa 
where $\eta^a$ is an asymptotic time-like Killing vector  and satisfies $\nabla_a\eta_b + \nabla_b \eta_a=0$ at $x^2\to+\infty$.
Since there are many asymptotic Killing vectors, we identify  a vector $\eta$ with another $\eta^\prime$
if there exist a vector $v_a=\eta_a-\eta^\prime_a$ which vanishes at $x^2\to+\infty$. 
Clearly $Q_{\rm Komar}[\eta]= Q_{\rm Komar}[\eta^\prime]$.
Under this identification, a collection of all independent asymptotic Killing vectors $\eta$ generate the isometry of the Minkowsiki spacetime, so that a number of independent vectors is $d(d+1)/2$ (translation and Lorentz transformation).
Thus the ADM energy is regarded as a conserved energy associated with the asymptotic time translation $\eta$ in the asymptotically flat spacetime. Since the ADM energy is (a half of) the Komar integral, we can write
\beqa
E_{\rm ADM} &=& {1\over 4\kappa} \int_{\Sigma_\infty} [d^{d-1}x]_a \,  \sqrt{-g}\nabla_b\left[ \nabla^{[a}\eta^{b]}\right],
\eeqa
where $\Sigma_\infty$ is a space-like surface whose boundary is given by $x^2\to +\infty$. 

\subsubsection{Asymptotically dS/AdS spacetime} 
As in the case of the asymptotically flat spacetime, we define the asymptotically deSitter(dS) or Anti-deSitter(AdS) spacetime
as the spacetime whose metric satisfies the vacuum Einstein equation with cosmological constant, $G_{ab} +\Lambda g_{ab} =0$
at $x^2\to\infty$. We then regard the isometry of the dS/AdS spacetime as a (representative of) asymptotic Killing vectors of this spacetime.   
The isometry of the dS is $SO(1,d)$, while that of the AdS is $SO(2,d)$.
Since it is possible to make the metric $g_{ab}$ static, the Killing vector $\eta$ for the time translation always exists.   
Thus the energy in these asymptotic spacetimes is defined using the asymptotic Killing vector  $\eta$ as
$ E^{\rm as}_{\rm dS/AdS} = Q_{\rm Komar}[\eta]$. 

\subsection{Cautions on charges from Noether's second theorem}
As we have already mentioned frequently,  the Noether's 2nd theorem tells that currents associated with local symmetries
are always conserved {\it without} using equations of motion of dynamical variable.
Thus conserved currents  and conserved charges do not reflect dynamical properties of the system.
Rather they are consequences of constraints \eqref{eq:constEOM} for Einstein gravity
among the quantities $E_{G}^{ab}$ and $E_\phi$, each of which would vanish at on-shell.    
Therefore it does not seem reasonable to define energy in general relativity by either pseudo-tensor or Komar integral
including the ADM energy or asymptotic charges.
Indeed Noether call the conservation law  from her 2nd theorem {\it improper}, referring statements by Hilbert and Klein\cite{Noether:1918zz}.

In addition,  both pseudo-tensor and Komar integral are easily modified by  an arbitrary total divergence term, which can be added without changing equations of motion,  so that they are not unique.
Furthermore, the pseudo-tensor depends on the choice of the coordinate as it is not covariant under general coordinate transformation. The Komar integral, on the other hand, is conserved for an arbitrary vector $\xi^a$, so that it may depend on a choice of $\xi^a$.

One may argue to define a physical Noether charge by regarding the local transformation restricted to constant parameters
as the ``global" transformation. However, this does not work except QED, since the conservation of the Noether's charge
associated with the  ``global" transformation is still a part of constraints implied by the local transformation.
QED is somewhat special, since the charge can be defined from the matter current, which is  U(1) gauge invariant. 

In the next section, we introduce our proposal for a proper and covariant definition of charges in general relativity, which are conserved only after equations of motion for gravity and matters are satisfied. 
We consider several examples  in order to compare our definition with those from the Noether's 2nd theorem.

 \section{Our physical definition vs. Noether's 2nd theorem in general relativity} 
 \label{sec:proposal}
 In this section, we first explain our recent proposal for the covariant definition of the energy and its generalization in general relativity\cite{Aoki:2020prb,Aoki:2020nzm}.
 We then compare our definition with those derived from the Noether's 2nd theorem in the previous section 
 for various examples with explicit calculations.
 
 \subsection{Our proposal for conserved non-Noether charge}
 We first summarize our proposal to define a conserved charge in general relativity\cite{Aoki:2020prb,Aoki:2020nzm}.
 We start with the Einstein equation given by
 \beqa
 G_{ab} + \Lambda g_{ab} &=& 2\kappa T_{ab},
 \label{eq:EE}
 \eeqa
 where the EMT $T_{ab}$ should be covariantly conserved, $ \nabla_a T^a{}_b = 0$, 
 as a  consequence of equations of motion for matters,
 since the left-hand side identically vanishes after applying $\nabla_a$
 due to the Bianchi identity.  
 
 We define a charge associated with a vector $\zeta^a$ as
 \beqa
 Q[\zeta] &=& \int_{\Sigma} [d^{d-1}x]_a\, \sqrt{-g}\, T^a{}_b \, \zeta^b,
 \label{eq:charge}
 \eeqa
 for a  space-like surface $\Sigma$.
 The  definition \eqref{eq:charge} is manifestly covariant under general coordinate transformations.
 With a similar argument  as discussed for $A^a{}_b$ around \eqref{eq:conv_law}, $Q[\zeta]$ is conserved ({\it i.e.} it does not depend on a choice of the space-like surface $\Sigma$),
 if  the standard conservation law $\partial_a   (\sqrt{-g}\, T^a{}_b \, \zeta^b) =0$ holds. 
 Therefore $\zeta^b$ must satisfy
  \beqa
  T^a{}_b (x)  \nabla_a \zeta^b (x) = 0
 \label{eq:zeta}
  \eeqa
  for $Q[\zeta]$ to be conserved,
  since  $\nabla_a T^{a}{}_b = 0$. 
  We call \eqref{eq:zeta} the conservation condition.
  
 Using  the conserved charge $Q[\zeta]$, we define the energy and its generalization in general relativity.
 There are three distinct cases for a choice of $\zeta$,
 which will be explained with explicit examples in the following subsections.
 We will also make comparisons with other definitions of the energy from the Noether's 2nd theorem in the previous section. 
   
 \subsection{Energy conservation by symmetry}
 If the metric, which is a solution to the Einstein equation \eqref{eq:EE}, is invariant under the time translation, then the (time-like) Killing vector $\xi^a$, defined by
 $ \nabla_a \xi_b +\nabla_b \xi_a=0$, exists.
 Since $T_{ab}=T_{ba}$, it is easy to see that $\zeta^a=\xi^a$ satisfies \eqref{eq:zeta}. 
 If the metric does not contain a time coordinate $x^0$, the Killing vector is given by $\xi^a = - \delta^a_0$ in such a coordinate.
 Thus 
 the conserved energy is defined by\cite{Aoki:2020prb}
 \beqa
 E&:=&Q(\zeta^a=-\delta^a_0) = -\int_{\Sigma} [d^{d-1}x]_a\, \sqrt{-g}\, T^a{}_0 = -\int_{\Sigma_0} [d^{d-1}x]_0 \sqrt{-g}\, T^0{}_0, 
 \label{eq:Energy}
 \eeqa 
 and the conservation is a consequence of the global time translational invariance of the on-shell metric, the solution to the Einstein equation, but is not a consequence implied by the local symmetry of the theory assumed in Noether's 2nd theorem.\footnote{While \eqref{eq:Energy} is not a Noether charge in the general relativity where $g_{ab}$ is dynamical,  this energy may be regarded as a conserve charge of the Noether's 1st theorem associated with the isometry for a {\it fixed background} metric. }
 In the 2nd equality, we present an expression for 
 a constant $x^0$ space-like surface $\Sigma_0$, where $ [d^{d-1}x^0]_0 := dx^1 dx^2\cdots dx^{d-1}$.
 
 \subsubsection{Vacuum energy}
 As a warmup, we consider a vacuum described by
 \beqa
 ds^2 = - f(r) (dx^0)^2 + {1\over f(r)} dr^2 +r^2d\Omega^2_{d-2} \quad f(r) = 1 - {2 \Lambda r^2\over (d-2)(d-1)}.
 \eeqa
As already mentioned,  the time-like Killing vector is given by $\xi^a = -\delta^a_0$, though it becomes space-like beyond the cosmological horizon
$r> r_H = \sqrt{\dfrac{(d-2)(d-1)}{2\Lambda}}$ for the positive cosmological constant $\Lambda$ (deSitter spacetime).
By definition, the energy of the vacuum is zero for our definition, $E_{\rm our}^{\rm vac} =0$, while energies from the Noether's 2nd theorem become
\beqa
E_{\rm pseudo}^{\rm vac} = E_{\rm Komar}^{\rm vac} = -{2\Lambda\Omega_{d-2}\over (d-2)\kappa}  \int r^{d-2} d r,   \quad
\Omega_{d-2} := {2\pi^{(d-1)/2}\over \Gamma({d-1\over 2})}.
\eeqa
Thus we have
\beqa
E_{\rm our}^{\rm vac} = E_{\rm pseudo}^{\rm vac} = E_{\rm Komar}^{\rm vac} = E_{\rm ADM}^{\rm vac} =0
\eeqa
for a flat spacetime, while
\beqa
E_{\rm our}^{\rm vac} =0, \quad
E_{\rm pseudo}^{\rm vac} = E_{\rm Komar}^{\rm vac} = E_{\rm dS/AdS}^{\rm vac} = -{2\Lambda\Omega_{d-2}\over (d-2)\kappa}  \int r^{d-2} d r
\to -\Lambda \times \infty
\eeqa
for non-zero cosmological constant, where the divergence comes from the divergent $r$ integral.
   
 \subsubsection{Schwarzschild black hole}
 \begin{figure}
\begin{center}
\includegraphics[width=0.7\textwidth,clip]{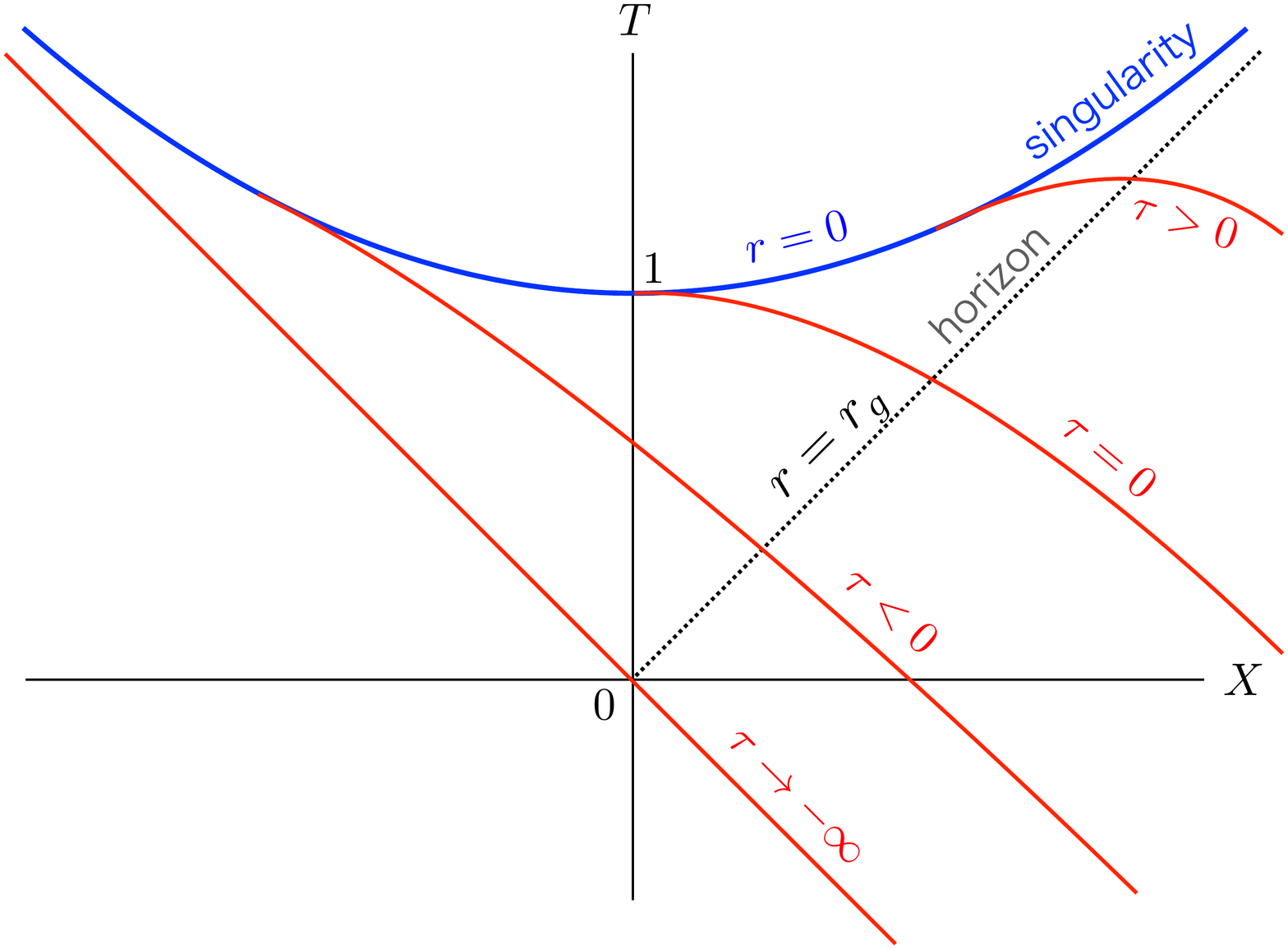}
\end{center}
 \label{fig:BH}
 \caption{The Schwarzschild black hole in the Kruskal-Szekeres like coordinate. A blue curve defined by $T=\sqrt{1+X^2}$ represents a black hole singularity at $r=0$ in the Eddington-Finkelstein  coordinates, while the dotted black line given by $T=X$ is the horizon as $r=r_g$. Red curves are constant $\tau$ surface in the Eddington-Finkelstein  coordinates at $\tau\to-\infty$, $\tau<0$, $\tau=0$ and $\tau>0$, respectively.  The physical region exists above the surface at $\tau\to-\infty$ and below the singularity surface at $r=0$. }
 \end{figure}
 As a non-trivial example, we consider the Schwarzschild black hole in $d$ dimensions, whose metric is given by
 \beqa
 ds^2 = -\left(1+u\right) d\tau^2 - 2u  d\tau dr +\left(1-u\right) dr^2 + r^2 d\Omega_{d-2}^2
  \eeqa 
in the Eddington-Finkelstein  coordinates, where 
\beqa
u &:=& \delta u -{2\Lambda r^2\over (d-2)(d-1)}, \quad
\delta u := - \left({r_g\over r}\right)^{d-3}, \quad r_g^{d-3} := 2 G M\theta(r), 
\eeqa
$r_g$ is the black hole horizon for a case with $\Lambda=0$, and $M$ is the mass of the black hole.
Here we introduce the step function $\theta(r)$ with $\theta(0)=0$ to  properly treat the singularity at $r=0$  in the distributional sense. Note that we can replace $\theta(r)$ with other  regularizations without changing discussions below\cite{Balasin:1993fn}.

The constant $\tau$ surface is normal to 
 \beqa 
 n_a = -\left(1-u\right)^{-1/2} \delta_a^\tau,  \qquad  n_a n^a=-1,
 \eeqa
 thus the constant $\tau$ surface is always space-like even inside the horizon except in the large $r$ region that $1-u < 0$ for the negative $\Lambda$ (AdS spacetime). 
 We illustrate the constant $\tau$ surface in the Kruskal-Szekeres like coordinates for $d=4$ and $\Lambda=0$ in Fig.~\ref{fig:BH}, where the metric becomes
 \beqa
 ds^2 &=& -{4 r_g^3 e^{-r/r_g}\over r} (dT^2 -dX^2) +r^2 d\Omega^2,\\
 X &=&e^{r\over 2 r_g}\left[\sinh\left({\tau\over 2 r_g}\right) + e^{-{\tau\over 2r_g}}{r\over 2r_g}\right], \
 T=e^{r\over 2 r_g}\left[\cosh\left({\tau\over 2 r_g}\right) - e^{-{\tau\over 2r_g}}{r\over 2r_g}\right],
 \eeqa
 and
 \beqa
\left.  {d T\over dX}\right\vert_\tau &=& {2r_g \cosh\left({\tau\over 2 r_g}\right) - e^{-{\tau\over 2r_g}}(r+2r_g)
 \over 2r_g \sinh\left({\tau\over 2 r_g}\right) + e^{-{\tau\over 2r_g}}(r+2r_g)}
  \eeqa
 for a fixed $\tau$.
 Toward the singularity ($r\to 0$), the coordinates behave as
 \beqa
 X&\to& \sinh\left({\tau\over 2 r_g}\right), \ T\to\cosh\left({\tau\over 2 r_g}\right), \
 \left.  {d T\over dX}\right\vert_\tau \to\tanh \left({\tau\over 2 r_g}\right),
  \eeqa
  while at horizon ($r=r_g$), they become
 \beqa
 X&=& T = {\sqrt{e}\over 2} e^{\tau\over 2 r_g}, \
 \left.  {d T\over dX}\right\vert_\tau = {2\sinh\left({\tau\over 2 r_g}\right) - e^{-{\tau\over 2 r_g}}
 \over 2 \cosh\left({\tau\over 2 r_g}\right) + e^{-{\tau\over 2r_g}}  },
  \eeqa  
and  at $r\to\infty$, they approach to
 \beqa
 X &\to&{r\over 2r_g} e^{(r-\tau)\over 2 r_g}, \
 T\to - {r\over 2r_g} e^{(r-\tau)\over 2 r_g}, \
  \left.  {d T\over dX}\right\vert_\tau \to -1.
 \eeqa
 
 The relevant component of the EMT is given by\cite{Aoki:2020prb} 
 \beqa
 T^\tau{}_\tau &=& {d-2\over 4\kappa }{\partial_r (r^{d-3}\delta u) \over r^{d-2}} =-{(d-2) M\over 8\pi}{\delta (r)\over r^{d-2}},
 \eeqa
 whose second expression agrees with the expression for the EMT by other regularizations in the distributional approach\cite{Balasin:1993fn}.
 Contrary to the general argument\cite{Geroch:1986jjl}, the EMT is well defined in the distributional sense, since it does not contain ill-defined products of two distributions. The energy is evaluated by the integral of this EMT over the  (d-1) dimensional constant $\tau$ surface (red curves in Fig.~\ref{fig:BH} for $d=4$) with the Killing vector\footnote{Although constant $\tau$ surfaces are space like, the Killing vector $\xi^\mu$ is time-like outside the horizon ( $r > r_g$) but space-like inside the horizon ($r<r_g$) for $\Lambda= 0$.
 In the case of non-zero cosmological constant,  the situation is similar but more complicated.  } $\xi^a=-\delta^a_\tau$ as
\beqa
E_{\rm our}^{\rm BH} &=& \int d^{d-1}x \sqrt{-g} T^\tau{}_a (-\delta^a_\tau) =  {(d-2) M\over 8\pi }\Omega_{d-2}
\int  dr \partial_r \theta(r)\nn \\
&=& {(d-2) \Omega_{d-2} \over 8\pi } M \left[\theta(\infty) -\theta(0)\right] 
= {(d-2) \Omega_{d-2} \over 8\pi } M,
\label{eq:BH_our}
\eeqa   
which exactly gives a mass of the black hole at $d=4$. While we here simply integrate $\partial_r\theta(r)$ over $r$, 
a direct use of $\delta(r)$ leads to the same result, showing a correctness of the distributional approach  as well as a famous relation  $\partial_r\theta(r)=\delta(r)$.

We now consider the black hole energies from the Noether's 2nd theorem.
Since we take the constant $\xi^a=-\delta^a_\tau$ in the case of the Schwarzschild black hole, the energy from the pseudo-tensor agrees with the Komar energy.
In addition, the result by the ``volume" integral with the delta function agrees with the one by the ``surface integral" without requiring a specific asymptotic behavior.  
Explicitly
\beqa
E_{\rm pseudo}^{\rm BH}&=& E_{\rm Komar}^{\rm BH}
= {1\over \kappa} \int d\Omega_{d-2} \int dr\, r^{d-2} R^\tau{}_\tau\xi^\tau
= {1\over 2\kappa} \int d\Omega_{d-2}\,  r^{d-2} \nabla^{[\tau}\xi^{r]}\nn \\
&=& \Omega_{d-2}\left[{(d-3) M\over 4\pi} - {2\lambda r^{d-1} \over \kappa (d-2)(d-1)}\right]
={(d-3) \Omega_{d-2} \over 4\pi} M + E_{\rm Komar}^{\rm vac}.
\eeqa
Thus $E_{\rm pseudo/dS/AdS}^{\rm BH}$ diverges for $\Lambda\not=0$, while
we can define the finite energy by subtracting the ``vacuum" contribution as
\beqa
\Delta E_{\rm 2nd}^{\rm BH} &:=& E_{\rm 2nd}^{\rm BH} - E_{\rm 2nd}^{\rm vac} =
{(d-3) \Omega_{d-2} \over 4\pi} M, 
\label{eq:BH_2nd}
\eeqa
where the word ``2nd"  represents the pseudo-tensor energy and the Komar energy including  the ADM energy and  the asymptotically dS/AdS energy.

If we compare \eqref{eq:BH_our} with \eqref{eq:BH_2nd}, we have
\beqa
{\Delta E_{\rm 2nd}^{\rm BH} \over E_{\rm our}^{\rm BH}} ={2 (d-3)\over d-2},
\eeqa
which becomes unity only at $d=4$.
Thus the covariant definition of the black hole energy in our proposal is in general different from ``energies" defined from the Noether's 2nd theorem, even after subtractions of the divergent vacuum contribution necessary for $\Lambda\not=0$,
though the difference appears only in the normalization.
A more distinct difference between the two definitions appears in the case of energies for a compact star\cite{Aoki:2020prb}. 

\subsection{Energy conservation without symmetry}
We next consider a case without Killing vector for the time translation.
Even in such a case where $\xi^a=-\delta^a_0$ is not a Killing vector anymore,
the energy defined by \eqref{eq:Energy} is time independent if the EMT and the metric satisfy 
\beqa
T^a{}_b \nabla_a \xi^b &=& -T^a{}_b \Gamma^b_{a 0} =0.
\label{eq:cond2}
\eeqa
In this case the energy $E$ is conserved but 
the conservation is NOT even a consequence of the global time translational invariance. 

\subsubsection{Gravitational collapse}
\begin{figure}[tbh]
\begin{center}
\includegraphics[width=1.0\hsize,angle=0]{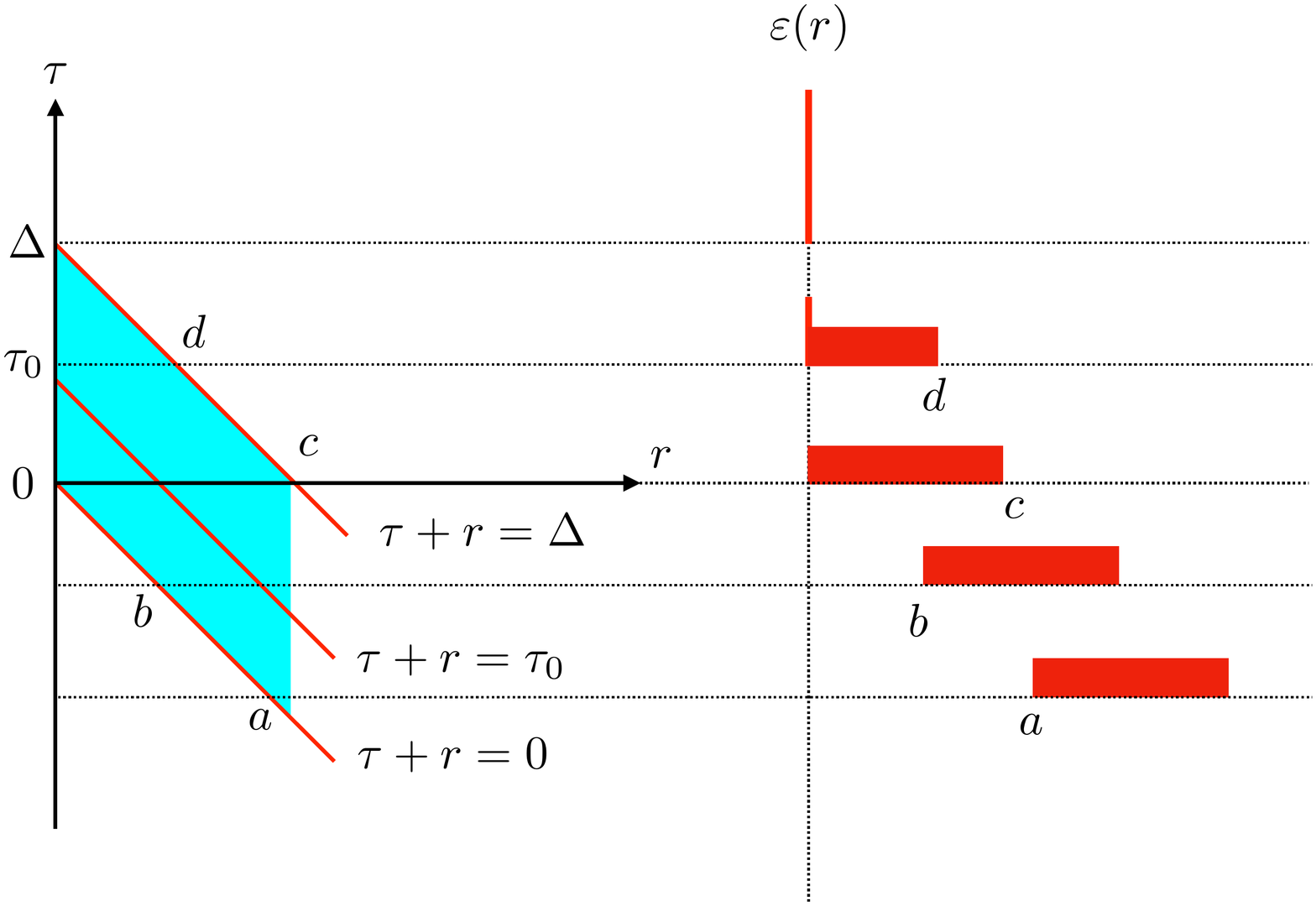}
\end{center}
\caption{(Left) Gravitational collapse of thick light shells in the Eddington-Finkelstein coordinate. Solids lines represent infalling lights which reach the origin at $\tau=0, \tau_0,\Delta$. (Right) The local energy density $\varepsilon(r)$ as a function of $r$
at various $\tau$. Here we consider $F(x)=x$ case as a simplest example, and 
$\varepsilon(r)$ has $\delta$ function contribution at $r=0$, represented by a thick vertical line.  
}
\label{fig:LightShell}
\end{figure}
Let us consider a simple model of gravitational collapses for thick light shells\cite{Adler:2005vn},
whose metric 
in the Eddington-Finkelstein coordinate is given by
\beqa
g_{ab} dx^a dx^b &=&- (1+u) d\tau^2 - 2u d\tau dr+(1-u)dr^2 + r^2 d\Omega_{d-2}^2,
\label{eq:g_collapse}
\eeqa
where $x^0=\tau$,  and 
\beqa
u(r,\tau) &:=& - \frac{m(r,\tau)}{r^{d-3}}-{2\Lambda r^2\over (d-2)(d-1)}, \\
m(r,\tau) &:=& \left\{
\begin{array}{lll}
2G M \theta(r),    &  \tau+r >  \Delta , & \mbox{I} \\
 2G M \theta(r) F\left(\dfrac{\tau+r}{\Delta}\right), & 0 \le    \tau+r \ge \Delta,  & \mbox{II} \\
 0, & \tau+r < \Delta,  & \mbox{III} \\  
\end{array}
\right. ,
\label{eq:shell}
\eeqa
where a monotonically increasing function $F(x)$ satisfies $F(0)=0$ and $F(1)=1$.
The vector $\xi^a=-\delta^a_0$ is NOT a Killing vector due to an existence of  the light shell  region (II),
while it becomes the Killing vector in Schwarzschild (I) and Minkowski (III) regions.
See Fig.~\ref{fig:LightShell} (Left), where solid lines represent infalling lights which reach the origin at $\tau=0, \tau_0, \Delta$. 

Since the metric in \eqref{eq:g_collapse} gives\cite{Aoki:2020nzm}
\beqa
\Gamma^0_{00}&=&{1+u\over 2} u_\tau -{u\over 2} u_r= - \Gamma^r_{r0}, \
\Gamma^0_{r0} ={u\over 2} u_\tau +{1-u\over 2} u_r, \
\Gamma^r_{00} =-{2+u\over 2} u_\tau +{1+u\over 2} u_r, \nn 
\eeqa
and 
\beqa
T^0{}_0 &=& \frac{(d-2)}{4\kappa} \frac{\left( r^{d-3} \delta u\right)_r}{r^{d-2}},\quad
T^r{}_r =  \frac{(d-2)}{4\kappa}\left[ \frac{ \left( r^{d-3}  \delta u\right)_r}{r^{d-2}}
-\frac{2 ( \delta u)_\tau}{r}\right], \nn \\
 T^0{}_r &=& \frac{(d-2)}{4\kappa} \frac{ (\delta u)_\tau}{r}= -  T^r{}_0,   \quad \delta u := - \frac{m(r,\tau)}{r^{d-3}},
\eeqa
the condition \eqref{eq:cond2} is satisfied for $\xi^a=-\delta^a_0$ as
\beqa
T^a{}_b \Gamma^b_{a 0} &=& (T^0{}_0 -T^r{}_r) \Gamma^0_{00} + T^0{}_r   (\Gamma^r_{00}
-\Gamma^0_{r0})= 0.
\eeqa

In this system, the energy \eqref{eq:Energy} is calculated as
\beqa
E(\tau) &=& -\int d^{d-1}x\, \sqrt{-g} T^0{}_0 ={(d-2) \Omega_{d-2}\over 16\pi G} \int_0^\infty dr\, \left[m(r,\tau)\right]_r.
\label{eq:Energy2}
\eeqa

For $\tau<0$ (before the collapse without a black hole),  \eqref{eq:Energy2} is evaluated as
\beqa
E(\tau) &=& ={(d-2) M \Omega_{d-2}\over 16\pi G}\int_{-\tau}^{\Delta-\tau}dr\, \partial_r (\theta F) = {(d-2) M \Omega_{d-2}\over 8\pi }:=E_{\rm tot}.
\eeqa

For $0\le \tau \le \Delta$ (during the collapse with a growing black hole), we obtain 
\beqa
E(\tau) &=& E_{\rm tot} \int_0^{\Delta-\tau} dr\, \partial_r(\theta F) = E_{\rm tot} \left[1 -\theta(0) F\left({\tau\over \Delta}\right)\right]
= E_{\rm tot} ,
\eeqa
which can be evaluated differently using $\partial_r(\theta F) =\delta(r) F + \partial_r F$ as
\beqa
E(\tau)&=& E_{\rm tot} \left[  F\left({\tau\over \Delta}\right)+\left\{ F(1) -F\left({\tau\over \Delta}\right)\right\}\right] = E_{\rm tot} ,
\eeqa
where the first term represents a mass of a growing black hole while the second one is an energy of remaining light shells. 

Finally for $\tau > \Delta$ (after the collapse with the final black hole), we evaluate the total energy as
\beqa
E(\tau) &=& E_{\rm tot}\int_0^\infty dr \delta(r) = E_{\rm tot},
\eeqa
which agrees with the mass of the final black hole.

The total energy is conserved as $E(\tau) = E_{\rm tot}$, and we plot typical distributions of the local energy density in Fig.~\ref{fig:LightShell} (Right). 

Other examples have also been discussed in \cite{Aoki:2020nzm}, and
gravitational collapses for more general energy-momentum tensors have been investigated recently in \cite{Yokoyama:2021nnw}.

\subsubsection{Comparison with energies in the Noether's 2nd theorem}
Since $\xi^a =-\delta^a_0$ is constant, 
the energy from the pseudo tensor and  the Komar energy agree.
We thus obtain
\beqa
E_{\rm pseudo} &=& E_{\rm Komar} = {\Omega_{d-2}\over 2\kappa}\int dr\, \partial_r \left[  r^{d-2}(u_r - u_\tau)\right]
= \left.  {\Omega_{d-2}\over 2\kappa} r^{d-2}(u_r - u_\tau)\right\vert_{r_0}^{r_1}
\eeqa
where $r_1=\Delta-\tau$ and $r_0=-\tau$ for $\tau < 0$, $r_1=\Delta-\tau$ and $r_0=0$ for $0<\tau < \Delta$, and
$u_\tau=0$ with $r_1=\infty$ and $r_0=0$ for $\tau > \Delta$.
We thus obtain  
\beqa
E_{\rm 2nd}&:=&E_{\rm pseudo} = E_{\rm Komar} =  {(d-3)\Omega_{d-2}\over 4\pi} M + E_{\rm Komar}^{\rm vac}
\eeqa
which again gives
\beqa
{E_{\rm 2nd} - E_{\rm 2nd}^{\rm vac}\over E_{\rm our}} = {2(d-3)\over d-4}.
\eeqa

\subsection{Conserved charge in the absence of energy conservation}
We finally consider the most general cases, where the Killing vector for time translation  is absent and \eqref{eq:cond2}
for the constant vector $\xi^a= -\delta^a_0$ is not satisfied. To define a conserved charge, which is a generalization of the energy, we must solve \eqref{eq:zeta}
for $\zeta^a(x) =\beta(x) n^a(x)$ with $n^a(x)=\frac{ d x^a(\eta)}{ d\eta}$ where $\eta$ is a parameter to characterize the time evolution of space-like surfaces $\Sigma_\eta$. 
(If we choose $\eta$ to be the global time $x^0$, we have $\zeta^a(x) = \beta(x) \delta^a_0$.)
As discussed in Ref.~\cite{Aoki:2020nzm},  a solution to \eqref{eq:zeta} always exists\footnote{The existence of such a vector field for a spherically symmetric gravitational system, known as the Kodama vector, was pointed out in Ref.~\cite{Kodama:1979vn}.} and is unique once  an initial condition for $\beta(x)$ is given at some $\eta=\eta_0$. 
Thus, using this $\zeta^a$, we can always define a conserved charge \eqref{eq:charge}, which is a generalization of the energy in general relativity. 

\subsubsection{Expanding universe}
As an example we consider a model of homogeneous and isotropic expanding universe in Einstein gravity with a cosmological constant $\Lambda$,  described by
the $d$-dimensional Friedmann-Lema\^itre-Robertson-Walker (FLRW) metric\cite{Friedmann:1924bb,Lemaitre:1931zza,Robertson:1935zz,Walker:1937aa}, 
\beqa
ds^2 &=& - (dx^0)^2 + a^2(x^0) \tilde g_{ij} dx^i dx^j,
\eeqa
where $a(x^0)$ is the scale factor dependent only on time $x^0$, and the $d-1$-dimensional Riemann tensor and
the Ricci tensor for $\tilde g_{ij}$ becomes 
\beqa
\tilde R_{ik}{}^{jl} = k\delta^j_{[i}\delta^l_{k]}, \quad \tilde R_i{}^j = k (d-2) \delta^i_j,
\eeqa
with
$k>=1$ (sphere), $0$ (flat space), $-1$ (hyperbolic space).

The EMT is given by the perfect fluid as
\beqa
T^0{}_0 &=& -\rho(x^0), \quad T^i{}_j = P(x^0)\delta^i_j, \quad T^0{}_j=T^i{}_0=0,
\eeqa
where $\nabla_a T^a{}_b = 0$ implies
\beqa
\dot\rho +(d-1)(\rho+P) {\dot a\over a} =0, \quad \dot\rho := \partial_0 \rho, \ \dot a:=\partial_0 a,
\eeqa
while the Einstein equation leads to
\beqa
8\pi G \rho={(d-1)(d-2)\over 2} {(k+\dot a^2)\over a^2} -\Lambda, \
8\pi G P = (2-d)\left[ {\ddot a\over a} +{(d-3)\over 2}{(k+\dot a^2)\over a^2}\right] +\Lambda.~~~~
\eeqa

In this case, the energy is given by
\beqa
E(x^0) &:=& -\int d^{d-1}x\, \sqrt{-g} T^0{}_0  = V_{d-1} a^{d-1}\rho, \quad 
V_{d-1}:= \int  d^{d-1}x\, \sqrt{\tilde g},
\eeqa
which is NOT conserved unless $P=0$, since
\beqa
{\dot E \over E}= -(d-1)  {\dot a\over a}{ P\over \rho} \not= 0.
\eeqa

To define a conserved charge as a generalization of energy, we take $\zeta^a = -\beta(x^0) \delta^a_0$ to satisfy
\eqref{eq:zeta}, which leads to\cite{Aoki:2020nzm} 
\beqa
-T^0{}_0 \dot\beta -T^i{}_j \Gamma^j_{i0} \beta &=& \rho \dot\beta - (d-1) P {\dot a\over a}\beta = 0,
\label{eq:beta}
\eeqa
where we use $\Gamma^j_{i0} = \dfrac{\dot a}{a}\delta^j_i $.
An existence of the second term violates the condition \eqref{eq:cond2} for the energy conservation.

A new conserved charge is thus given by
\beqa
S(x^0) &:=& \int d^{d-1}x\, \sqrt{-g} (- T^0{}_0)\beta =V_{d-1} a^{d-1} \rho \beta,
\eeqa
which is manifestly conserved as
\beqa
{\dot S\over S} &=& {\dot E\over E}  +  {\dot \beta\over \beta} =  -(d-1)  {\dot a\over a}{ P\over \rho} + (d-1) {P\over \rho} {\dot a\over a}
=0.
\eeqa
The energy non-conservation is compensated by the second term.

What is this conserved charge $S$ ?
If we define densities  $e(x^0):= E(x)/V_{d-1}=\rho(x)v(x^0)$ and $s(x^0):=S(x)/V_{d-1}=e(x^0)\beta(x^0)$,
where
$v(x^0) := a(x^0)^{d-1}$ is a local volume element at time $x^0$,
we obtain
\beqa
{d s\over dx^0} =  {d e\over dx^0} \beta + e {d \beta\over dx^0} = \left({d e\over dx^0} + P {d v\over dx^0} \right) \beta,
\eeqa
where we use \eqref{eq:beta}.
This relation is very similar to the first law of thermodynamics as
\beqa
T ds = d e + P dv, 
\eeqa 
if we identify $\beta=\dfrac{1}{T}$ as an inverse temperature.
We thus interpret $S$ as the total entropy of the universe, which is conserved in
the FLRW universe\cite{Aoki:2020nzm}
\footnote{Without a mixing between time and space components for the metric and the energy-momentum tensor, the entropy density $s$ is also conserved\cite{Aoki:2020nzm}. }.
 In addition, $\beta(x^0)$ is regarded as the time-dependent inverse temperature of the universe. 
It is easy to see that the temperature decreases as the universe expands, since
\beqa
{\dot\beta\over \beta}=(d-1){P \dot a\over \rho a} > 0.
\eeqa

Even in more general cases, the entropy $S$ so defined is conserved in general relativity\cite{Aoki:2020nzm}.

\vskip 0.5cm

Although we assume the Einstein equation \eqref{eq:EE} for analyses in this section, our definition
of the conserved charge \eqref{eq:charge} works for an arbitrary theory of general relativity whose equation of motion
is given by $\tilde G_{ab}  = 2\kappa T_{ab}$ instead of \eqref{eq:EE} , where  $\tilde G_{ab}  $ is an arbitrary 2nd rank symmetric tensor composed of the metric $g_{ab}$ which satisfies $\nabla_a \tilde G^{a}{}_b =0$. 

\subsubsection{Conserved charge from the 2nd theorem}
Let us consider the conserved charge from the Noether's 2nd theorem for the FLRW universe.
In the case of the pseudo-tensor, we have
\beqa
A^0{}_0 &=& {\sqrt{-g}\over 2\kappa} \left[ 2R^0{}_0 + g^{0c} \Gamma^d_{d0,c} - g^{cd} \Gamma^0_{cd,0}\right] =0,
\eeqa
where we use
\beqa
R^0{}_0 &=& (d-1) {\ddot{a}\over a}, \quad
\Gamma^0_{ij} =a \dot a \tilde g_{ij}.
\eeqa
Thus $E_{\rm pseudo}^{\rm FLRW} =0$, which is conserved but physically  trivial.

The conserved current density for the Komar energy is given by
\beqa
J^a[\xi] = {1\over 2\kappa} \sqrt{-g} \nabla_b \left[\nabla^{[a}\xi^{b]}\right], 
\eeqa
where we take a non-constant $\xi^a=\gamma(x^0,r)\delta^a_0$. Here the $d-1$ dimensional metric is parametrized as 
\beqa
\tilde g_{ij}dx^i dx^j  = {dr^2\over 1-kr^2} + r^2 h_{kl} dx^k dx^l
\eeqa
with the $d-2$ dimensional metric $h_{kl}$ for a unit sphere.
Since $r=0$ is not a special point in the $d-1$ dimensional space, $\gamma(x^0,r=0)$ must be finite. 
Non-zero components of the current density with this choice of $\xi^a$ become
\beqa
J^0(x) &=& -{a^{d-3}\sqrt{h} \over 2\kappa}\partial_r ( r^{d-2}\sqrt{1-k r^2} \partial_r \gamma),  \
J^r(x) = { r^{d-2} \sqrt{1-k r^2} \sqrt{h}\over 2\kappa} \partial_0 ( a^{d-3} \partial_r \gamma), ~~~~~~
\eeqa
where $h$ is the determinant of $h_{kl}$. For the conservation of the Komar energy, the boundary contribution at $r\to r_\infty$, 
where $r_\infty=\infty$ for $k\le 0$ or $r_\infty^2=1/k$ for $k>0$,
given by 
\beqa
\lim_{r\to r_\infty} \int_{x^0_i}^{x^0_f}dx^0 \int d^{d-2}x\, J^r(x) &=&\left.
\lim_{r\to r_\infty}{\Omega_{d-2} \over 2\kappa} a^{d-3}(x^0) r^{d-2} \sqrt{1-k r^2} \partial_r \gamma(x^0,r) \right\vert_{x^0=x^0_i}^{x^0=x^0_f},~~~~~~
\eeqa 
must vanish,\footnote{If the space is a $(d-1)$-sphere ($k>0$), there should be no need for spatial boundary condition.  Using the spherical coordinate and polar angle $\theta$ to set $r$ as $\sqrt{k}r=\sin\theta$,  the boundary condition \eqref{eq:bc} reads
$\displaystyle{\lim_{\theta\rightarrow\pi}}\left(\sin\theta\right)^{d-2} \partial_\theta \gamma = 0$. It is obvious that this equation is trivially satisfied.
} 
where $\Omega_{d-2} := \int d^{d-2} x \sqrt{h}$ is the volume of the $d-2$ dimensional unit sphere. 
Thus $\gamma(x^0,r)$ must satisfy\beqa
\lim_{r\to r_\infty} r^{d-2} \sqrt{1-k r^2} \partial_r \gamma(x^0,r) =0.
\label{eq:bc}
\eeqa
Under this condition, the Komar energy is evaluated as
\beqa
E_{\rm Komar}^{\rm FLRW} &=& \int d^{d-1} J^0(x) =\left. -{\Omega_{d-2}\over 2\kappa} a^{d-3}(x^0) r^{d-2}\sqrt{1-k r^2}  \partial_r \gamma(x^0,r)\right\vert_{r=0}^{r=r_\infty} =0. \eeqa
Thus, the Komar energy is conserved but physically  trivial, as in the case of the pseudo-tensor. 

\subsection{Initial condition of $\zeta^a(x) = \beta(x) n^a(x)$ }
As mentioned before,  \eqref{eq:zeta} has a unique solution if the initial condition for $\beta(x)$ is given.
A  priori, there is no principle for a choice of the initial $\beta(x)$.  Since $\beta(x)$ physically represents a local inverse temperature,
we have to determine a local temperature distribution of matters from the matter energy momentum tensor $T^a{}_b(x)$ at some $x^0$
in order to fix the initial value of $\beta(x)$.
In the case of the FLRW universe, since matters are uniformly distributed,  it is natural to take the initial $\beta(x)$ to be uniform as well.
For general cases, however,
 it has not been known to define the local temperature from matter distributions. 
We leave this important problem to future investigations.

 \section{Conclusion and discussion}
 \label{sec:concl}
 In this paper, we  have shown that the pseudo-tensor as well as the Komar integral types of the energy including their quasi-local expressions are inappropriate to give the physically meaningful definition of the energy in general relativity. This is because their conservation  derived from the Noether's 2nd theorem
 is merely an identity representing a constraint by the local invariance rather than a consequence of the dynamics.
 The Noether's 2nd theorem covers almost all existing definitions of the energy in general relativity including the
 Abbott-Deser definition\cite{Abbott:1981ff} in addition to others mentioned in the main text.
 
In contrast, our proposal utilizes equations of motion to derive the conservation of the energy/entropy without using the Noether's theorem.
Thus, more than 100 years after Einstein's proposal,
our definition finally provides a proper and covariant definition of the energy whose generalization as the entropy is always 
conserved in general relativity.

The form of the conserved entropy in general relativity depends explicitly on the on-shell $g_{ab}$, the solution to the Einstein equation, through $\zeta^a(x)=\beta(x) n^a(x)$ in \eqref{eq:zeta}, where $\beta(x)$ is determined {\it after} the Einstein equation is solved. Thus we cannot predict how the spacetime evolves in time using the conservation law of the entropy, unlike 
the standard conservation law of the energy in the flat space time, which often gives manifest constraints to dynamics of the system.  

As evident from the form of the conserved current,  $J^a(x):= T^a{}_b(x) n^b(x) \beta(x)$,
the energy/entropy in general relativity is carried only by the matter energy momentum tensor.
This means that  gravitational fields including (Ricci flat) gravitational waves cannot carry the energy/entropy in general relativity.
Even though one may invent another definition of a conserved energy for gravitational fields,
it is still true that there exists  the conserved energy/entropy carried only by matters in general relativity.
Thus, it is interesting to reanalyze the binary star merger in terms of the conserved entropy, since
it has been interpreted that the energy loss through the emission of gravitational waves from rotating binary stars  causes their merger.
Last but not least, a fact that gravitational fields carry no energy/entropy give a very strong constraint to 
a theory of quantum gravity if it indeed exists.
For example, although a graviton, a quanta of the quantized gravity, carries the energy/entropy, 
a quantum average of an energy/entropy exchange between matter and gravity field must vanishes in the classical limit ($\hbar\to 0$).

\section*{Acknowledgements}
We would like to thank Drs.~Kohei Kamada, Satoshi Iso, Taichiro Kugo, Seiji Terashima, Shuichi Yokoyama and Tamiaki Yoneya for useful discussions.
 This work is supported in part by the Grant-in-Aid of the Japanese Ministry of Education, Sciences and Technology, Sports and Culture (MEXT) for Scientific Research (Nos.~JP16H03978, JP18K03620, JP18H05236.). 
 
 \appendix
 \section{Noether's 2nd theorem}
 \label{sec:Noether2nd}
For the sake of readers, we give a derivation of the Noether's 2nd theorem\cite{Noether:1918zz}.
In the case of general relativity,  see also an appendix of \cite{Utiyama:1984bc}, which however seems to be not  recognized well in
the community. 

 \subsection{Invariant variational theory}
 Let us consider an integral of Lagrangian $L$ over an arbitrary $d$-dimensional  region $\Omega$  given by
 \beqa
 S_\Omega &=& \int_\Omega d^d x\, L(\varphi_n,\varphi_{n,\mu} , \varphi_{n,\mu\nu}),
 \eeqa
 where $\varphi_{n,\mu} := \partial_\mu \varphi_n$,  
 $\varphi_{n,\mu\nu} := \partial_\nu \partial_\mu\varphi_n$,
 and $n=1,2,\cdots, N$ labels $N$ different fields.
 Unlike the standard Lagrangian which, contains at most the first derivatives of  $\varphi$,
the above  $L$ also contains the second derivatives of $\varphi_n$, which are necessary for the Einstein's general relativity.
Our discussion below can be extended to a more general $L$ including derivatives of $\varphi_n$ higher than the second, though  the formula becomes more complicated.
 
 A variation of  $S_\Omega$ is evaluated as
 \beqa
 \delta_v S &=& \int_\Omega d^dx\,\left[
 \frac{\partial L}{\partial \varphi_n} \delta_v \varphi_n +  \frac{\partial L}{\partial \varphi_{n,\mu}} \partial_\mu \delta_v\varphi_n +  \frac{\partial L}{\partial \varphi_{n,\mu\nu}}  \partial_\nu\partial_\mu \delta_v\varphi_n 
 \right] 
 \nn \\
 &=&  \int_\Omega d^dx\,\left\{ [L]^n \delta_v\varphi_n + \partial_\mu \Theta^\mu\left(  \delta_v\varphi_n \right)\right\}, 
 \eeqa
 where 
 \beqa
 [L]^n &:=&  \frac{\partial L}{\partial \varphi_n}  -\partial_\mu   \frac{\partial L}{\partial \varphi_{n,\mu}} 
 +\partial_\nu \partial_\mu   \frac{\partial L}{\partial \varphi_{n,\mu\nu}}  , \\
 \Theta^\mu(\delta_v\varphi_n)&=&  \left(  \frac{\partial L}{\partial \varphi_{n,\mu}}  -\partial_\nu \frac{\partial L}{\partial \varphi_{n,\mu\nu}} \right)\delta_v\varphi_n  +  \frac{\partial L}{\partial \varphi_{n,\mu\nu}}\partial_\nu \delta_v\varphi_n .
 \eeqa
 If we take an arbitrary variation of $\varphi_n$ such that $\delta_v\varphi_n = \partial_\mu\delta_v\varphi_n=\partial_\mu\partial_\nu\delta_v\varphi_n=0$ on the  boundary of $\Omega$,  the total divergent term, $\partial_\mu \Theta^\mu$, vanishes.  
 Thus  $\delta_v S_\Omega=0$ for an arbitrary variation  of $\varphi_n$ under this constraint 
 implies $[L]^n=0$, which gives equations of motion for $ \varphi_n$ .
 
 In addition we assume that $S_\Omega$ is invariant under the following transformation,
 \beqa
 x^\mu &\rightarrow& (x^\prime)^\mu = f^\mu(x) , \quad
 \varphi_n(x) \rightarrow \varphi^\prime_n(x^\prime) = F_n(\varphi, x), 
 \eeqa
 whose infinitesimal version is given by
 \beqa
 (x^\prime)^\mu &=& x^\mu +\delta x^\mu, \quad
 \varphi^\prime_n(x^\prime) = \varphi_n(x) +\delta \varphi_n(x) .
 \eeqa
 Note that $\delta$ can be  global as well as local transformations, but is
 different from the variation $\delta_v$ to derive equations of motion for $\varphi_n$.
 Since
\beqa
\frac{\partial (x^\prime)^\mu}{\partial x^\nu} =\delta^\mu_\nu +\frac{\partial \delta x^\mu}{\partial x^\nu}, \quad
 \Rightarrow \quad
\frac{\partial x^\nu}{\partial (x^\prime)^\mu} =\delta_\mu^\nu -\frac{\partial \delta x^\nu}{\partial x^\mu}, \
\eeqa
we obtain
\beqa
\delta ( \partial_\mu F(x) ) &:=& {\partial F^\prime(x^\prime) \over\partial (x^\prime)^\mu} -  {\partial F(x) \over \partial x^\mu}
= \partial_\mu \delta F(x) - \partial_\nu F(x) \partial_\mu \delta x^\nu,
\eeqa
 where $\delta F(x) := F^\prime(x^\prime) - F(x)$. This shows that  $\delta$ does not commute with the derivative $\partial_\mu$
 due to the second term.
 We thus introduce another variation $\bar\delta F(x) := F^\prime(x) - F(x)$, which commutes with derivatives as
 \beqa
 \bar\delta (\partial_\mu F) = \partial_\mu \bar\delta F, \qquad
 \delta F =\bar\delta F +\partial_\mu F \delta x^\mu.
 \eeqa

 Then the variation of $S_\Omega$ under $\delta$ is evaluated as
 \beqa
 \delta S_\Omega &=& \int_\Omega d^dx\, \left[
  \frac{\partial L}{\partial \varphi_n} \delta \varphi_n + \frac{\partial L}{\partial \varphi_{n,\mu}} \delta ( \varphi_{n,\mu}) + \frac{\partial L}{\partial \varphi_{n,\mu\nu}} \delta ( \varphi_{n,\mu\nu})
  + L \partial_\mu \delta x^\mu
 \right] \nn \\
& =& \int_\Omega d^dx\, \left[ [L]^n \bar\delta\varphi_n + \partial_\mu \left\{ \Theta^\mu(\bar\delta \varphi_n) + L \delta x^\mu\right\}
\right] 
\nn \\
&=& \int_\Omega d^dx\, \left[ [L]^n (\delta\varphi_n-\varphi_{n,\mu}\delta x^\mu) + \partial_\mu \left\{\Theta^\mu( \delta \varphi_n) - E^\mu{}_\nu \delta x^\nu
-G^{\mu\alpha}{}_\nu \partial_\alpha \delta x^\nu\right\}
\right] \label{eq:dS2}
 =0,~~~~~~~
 \eeqa
 where we use $d^dx^\prime = (1+ \partial_\mu \delta x^\mu) d^dx$ and 
 \beqa
 \partial_\alpha L(\varphi_n, \varphi_{n,\mu}, \varphi_{n,\mu\nu}) &=&   \frac{\partial L}{\partial \varphi_n} \partial_\alpha \varphi_n + \frac{\partial L}{\partial \varphi_{n,\mu}} 
    \partial_\alpha \varphi_{n,\mu} + \frac{\partial L}{\partial \varphi_{n,\mu\nu}} 
    \partial_\alpha \varphi_{n,\mu\nu}, 
\eeqa
and we define
\beqa    
    E^\mu{}_\nu &:=& \frac{\partial L}{\partial \varphi_{n,\mu}} \varphi_{n,\nu} 
    -\partial_\alpha \frac{\partial L}{\partial \varphi_{n,\mu \alpha}} \varphi_{n.\nu}
    +\frac{\partial L}{\partial \varphi_{n,\mu \alpha}} \varphi_{n,\nu\alpha}
    -\delta^\mu_\nu L , \\
    G^{\mu\alpha}{}_\nu &:=&\frac{\partial L}{\partial \varphi_{n,\mu \alpha}}\varphi_{n,\nu} .
 \eeqa
 
 \subsection{Noether's 1st theorem}
 Before considering the Noether's 2nd theorem,  we derive the well-known Noether's 1st theorem from the invariant variational theory.
 If we take
 \beqa
 \delta x^\mu= \epsilon^r f^\mu_r(x), \quad \delta\varphi_n =  \epsilon^r F_{r,n}(x,\varphi),
 \eeqa
  where $\epsilon^r$ ($r=1,2,\cdots,R$ ) are arbitrary constant parameters while $f_r^\mu(x)$ and $F_{r,n}(x,\varphi)$ are given functions of arguments.
  Then (\ref{eq:dS2}) becomes
  \beqa
  \delta S_\Omega &=& \epsilon^r \int_\Omega d^dx\, \left\{ [L]^n X_{n,r} + \partial_\mu J^\mu{}_r \right\} = 0,
  \eeqa
 where 
 \beqa
 X_{n,r} &:=& F_{r,n} -\varphi_{n,\mu} f^\mu_r, \nn \\
 J^\mu{}_r  &:=&  \left( \frac{\partial L}{\partial \varphi_{n,\mu}} 
 -\partial_\alpha \frac{\partial L}{\partial \varphi_{n,\mu\alpha}} \right) F_{r,n}
 - E^\mu{}_\nu f_r^\nu 
+ \frac{\partial L}{\partial \varphi_{n,\mu\nu}} \partial_\nu F_{r.n} 
-G^{\mu\alpha}{}_\nu \partial_\alpha f^\nu_r,
 \eeqa
 and summations over repeated indices including $n$ are understood.
 
 Since we can take $\Omega$ arbitrarily small, we obtain
 \beqa
 [L]^n X_{n,r} + \partial_\mu J^\mu{}_r &=& 0.
 \eeqa
 Thus, if equation of motions are satisfied as $[L]^n=0$ for ${}^\forall n$, 
 there appear $R$ conserved currents $J^\mu{}_r$ such that $\partial_\mu J^\mu{}_r =0$, as a consequence of the global symmetry   generated by parameters $\epsilon^r$. This is the famous Noerher's 1st theorem.

 \subsection{Noether's 2nd theorem}
 Let us consider the local transformation generated by $\xi^r(x)$ as
 \beqa
 \delta x^\mu &=& \xi^r f_r^\mu(x), \quad
 \delta\varphi_n =\xi^r F_{r,n}(x,\varphi) + \xi^r_{,\mu} F_r{}^\mu{}_{,n}(x,\varphi),
 \label{eq:tranL}
 \eeqa
 where $r=1,2,\cdots, R$ labels $R$ different generators,  and we denote
 $\xi^r_{,\mu} := \partial_\mu\xi^r$, $\xi^r_{,\mu\nu} :=\partial_\nu \partial_\mu\xi^r$ and so on, as before.
 Then eq.~(\ref{eq:dS2}) becomes
 \beqa
 &&\int_\Omega d^d x \Bigl[ \xi^r\Bigl\{ [L]^n \left( F_{r,n} - \varphi_{n,\mu} f_r^\mu\right)
 -\partial_\mu \left( [L]^n F_r{}^\mu{}_{,n} \right)
 \Bigr\}
 +\partial_\mu\left( A^\mu{}_r \xi^r + B^{\mu,\nu}{}_r \xi^r_{,\nu}
 +C^{\mu,\nu\alpha}{}_r \xi^r_{,\nu\alpha}
 \right) \Bigr]\nn \\
 & =& 0, 
 \label{eq:N2nd}
 \eeqa
 where
 \beqa
 A^\mu{}_r &:=& \left( \frac{\partial L}{\partial \varphi_{n,\mu}} 
 -\partial_\alpha \frac{\partial L}{\partial \varphi_{n,\mu\alpha}} \right) F_{r,n}
 - E^\mu{}_\nu f_r^\nu +[L]^n F_r{}^\mu{}_{,n}
+ \frac{\partial L}{\partial \varphi_{n,\mu\nu}} \partial_\nu F_{r.n} 
-G^{\mu\alpha}{}_\nu \partial_\alpha f^\nu_r
 , \nn \\
 B^{\mu,\nu}{}_r &:=&  \left( \frac{\partial L}{\partial \varphi_{n,\mu}} 
 -\partial_\alpha \frac{\partial L}{\partial \varphi_{n,\mu\alpha}} \right) 
   F_r{}^\nu{}_{,n}  +\frac{\partial L}{\partial \varphi_{n,\mu\nu}}  F_{r.n} 
  + \frac{\partial L}{\partial \varphi_{n,\mu\alpha}} \partial_\alpha F_{r}{}^\nu{}_{n} 
-G^{\mu\nu}{}_\alpha f^\alpha_r, \nn \\
C^{\mu,\nu\alpha}{}_r &:=&  \frac{\partial L}{\partial \varphi_{n,\mu\nu}} F_r{}^\alpha{}_n = C^{\nu,\mu\alpha}{}_r,
 \eeqa
 and summations over repeated indices including $n$ are also understood.
   
 As before  we can take $\Omega$ arbitrarily small.
 In addition, as opposed to the case of the global symmetry,
 we can also take $\xi^r=\xi^r_{,\mu}=\xi^r_{,\mu\nu}=0$ on $\partial \Omega$ (the boundary of $\Omega$).
 This choice leads to
 \beqa
 [L]^n \left( F_{r,n} - \varphi_{n,\mu} f_r^\mu\right)
 -\partial_\mu \left( [L]^n F_r{}^\mu{}_{,n} \right) = 0,
 \eeqa
 which can give  $R$ constraints on $N$ equation of motions.
 Putting this back into (\ref{eq:N2nd}) with an arbitrary $\Omega$ and $\xi$, we obtain
 \beqa
 \partial_\mu\left( A^\mu{}_r \xi^r + B^{\mu,\nu}{}_r \xi^r_{,\nu}
 +C^{\mu,\nu\alpha}{}_r \xi^r_{,\nu\alpha} 
 \right) = 0,
 \label{eq:conv2nd}
 \eeqa
 which reduces to
  \beqa
 \partial_\mu (A^\mu{}_r) \xi^r &+& \left( A^\nu{}_r  + \partial_\mu B^{\mu,\nu}{}_r \right) \xi^r_{,\nu}
 + \frac{1}{2} \left(B^{\mu,\nu}{}_r + B^{\nu,\mu}{}_r
 + 2\partial_\alpha  C^{\alpha,\mu\nu}{}_r \right) \xi^r_{,\nu\mu} 
  \nn \\
 &+& \frac{1}{3}\left( C^{\mu,\nu\alpha}{}_r +C^{\nu,\alpha\mu}{}_r +C^{\alpha,\mu\nu}{}_r \right)
 \xi^r_{,\nu\alpha\mu}  
= 0.
 \label{eq:conv2ndB}
 \eeqa
 Since $\xi^r$, $\xi^r_{,\nu}$, $\xi^r_{,\mu\nu}$ and $\xi^r_{,\mu\nu\alpha}$ in \eqref{eq:conv2ndB} are all arbitrary,
 we can conclude
 \beqa
 \partial_\mu A^\mu{}_r &=&  0, \nn \\
 A^\nu{}_r  + \partial_\mu B^{\mu,\nu}{}_r &=& 0, \nn \\
 B^{\mu,\nu}{}_r + B^{\nu,\mu}{}_r + 2\partial_\alpha  C^{\alpha,\mu\nu}{}_r &=& 0, \nn \\
 C^{\mu,\nu\alpha}{}_r +C^{\nu,\alpha\mu}{}_r +C^{\alpha,\mu\nu}{}_r &=& 0,
 \label{eq:BCD}
 \eeqa
 as constraints for {\it off-shell} $\varphi_n$.
Thus the constraints are expressed by the form of conservation as
\beqa
\partial_\mu J^\mu{}_r=0, \quad r=1,2,\cdots, R,
\eeqa
where
 \beqa
 J^\mu{}_r &:=& A^\mu{}_r = - \partial_\nu \tilde B^{\nu,\mu}{}_r , \quad
 \tilde B^{\nu,\mu}{}_r := {1\over 2} B^{[\nu,\mu]}{}_r  - {1\over 3}\partial_\alpha  C^{[\nu,\mu]\alpha}{}_r .
 \eeqa
 These constraints that $\partial_\mu J^\mu{}_r=0$, however, are not invariant under \eqref{eq:tranL} due to a presence of uncontracted  index $r$.

 \eqref{eq:conv2nd} is also regarded as a conservation equation that
 \beqa
 \partial_\mu J^\mu{}[\xi] = 0,
 \eeqa
 where $J^\mu[\xi]$ is defined as
 \beqa
 J^\mu [\xi] &=&  A^\mu{}_r \xi^r + B^{\mu,\nu}{}_r \xi^r_{,\nu}+C^{\mu,\nu\alpha}{}_r \xi^r_{,\nu\alpha} \label{eq:WK}
  \eeqa
 This conservation equation is manifestly invariant under   \eqref{eq:tranL}, since uncontracted indices are absent.
Using \eqref{eq:BCD} one can further rewrite $J^\mu(x)$ as
 \beqa
 J^\mu [\xi]
  &=& -(\partial_\nu B^{\nu,\mu}{}_r) \xi^r + B^{\mu,\nu}{}_r \xi^r{}_{,\nu}
 + C^{\mu,\nu\alpha}{}_r\xi^r{}_{,\nu\alpha}
 = -\partial_\nu (B^{\nu,\mu}{}_r \xi^r) +B^{\{\mu,\nu\}}{}_r \xi^r{}_{,\nu}+ C^{\mu,\nu\alpha}{}_r\xi^r{}_{,\nu\alpha}\nn \\
 &=&-\partial_\nu( B^{\nu,\mu}{}_r\xi^r + 2 C^{\nu,\mu\alpha}{}_r \xi^r{}_{,\alpha})
 +(2C^{\alpha,\mu\nu}{}_r + C^{\mu,\nu\alpha}{}_r) \xi^r{}_{,\nu\alpha}\nn\\
& =& -\partial_\nu( B^{\nu,\mu}{}_r\xi^r + 2 C^{\nu,\mu\alpha}{}_r \xi^r{}_{,\alpha}).
 \eeqa
Thus the current $J^\mu[\xi]$ turns out to be a total divergence.
\vskip 0.5cm 
 
 Let us remind readers that equations of motion are not employed to derive the conservation equations in the Noether's 2nd theorem. 
 Even if we restrict $\xi^\mu_r(x)$ to a constant as $\xi^\mu_r(x) =\epsilon^\mu_r$, 
 {\it off-shell} conservation equations still hold, so that
 conservations cannot be regarded as the dynamical ones in the standard Noether's 1st theorem. 
 Noether herself (as a word by Hilbert and Klein) called such conservations {\it improper}\cite{Noether:1918zz}
 and  distinguished them from {\it proper} conservations in  the 1st theorem.


\begin{thebibliography}{99}
\bibitem{Einstein:1916}
A.~Einstein,
{\it Ann. der. Phys. Ser.}4, {\bf 49} (1916), pp. 769-822

\bibitem{Arnowitt:1962hi}
R.~L.~Arnowitt, S.~Deser and C.~W.~Misner,
in {\it Gravitaion: an introduction to current research}, L. Witten, ed. (Wiley, New York, 1962).
See also 
Gen. Rel. Grav. \textbf{40}, 1997-2027 (2008)
doi:10.1007/s10714-008-0661-1.

\bibitem{Bondi:1962px}
H.~Bondi, M.~van der Burg and A.~Metzner,
Proc. Roy. Soc. Lond. A \textbf{A269}, 21-52 (1962)
doi:10.1098/rspa.1962.0161.

\bibitem{Brown:1992br}
J.~D.~Brown and J.~W.~York, Jr.,
Phys. Rev. D \textbf{47} (1993), 1407-1419
doi:10.1103/PhysRevD.47.1407
[arXiv:gr-qc/9209012 [gr-qc]].

\bibitem{Hawking:1995fd}
S.~Hawking and G.~T.~Horowitz,
Class. Quant. Grav. \textbf{13}, 1487-1498 (1996)
doi:10.1088/0264-9381/13/6/017
[arXiv:gr-qc/9501014 [gr-qc]].

\bibitem{Horowitz:1998ha}
G.~T.~Horowitz and R.~C.~Myers,
Phys. Rev. D \textbf{59}, 026005 (1998)
doi:10.1103/PhysRevD.59.026005
[arXiv:hep-th/9808079 [hep-th]].

\bibitem{Balasubramanian:1999re}
V.~Balasubramanian and P.~Kraus,
Commun. Math. Phys. \textbf{208}, 413-428 (1999)
doi:10.1007/s002200050764
[arXiv:hep-th/9902121 [hep-th]].

\bibitem{Ashtekar:1999jx}
A.~Ashtekar and S.~Das,
Class. Quant. Grav. \textbf{17}, L17-L30 (2000)
doi:10.1088/0264-9381/17/2/101
[arXiv:hep-th/9911230 [hep-th]].

\bibitem{DeHaro:2021gdv}
S.~De Haro,
[arXiv:2103.17160 [physics.hist-ph]].

\bibitem{Aoki:2020prb}
S.~Aoki, T.~Onogi and S.~Yokoyama,
Int. J. Mod. Phys. A \textbf{36} (2021) no.10, 2150098
doi:10.1142/S0217751X21500986
[arXiv:2005.13233 [gr-qc]].

\bibitem{Aoki:2020nzm}
S.~Aoki, T.~Onogi and S.~Yokoyama,
Int. J. Mod. Phys. A \textbf{36} (2021) no., 2150201
doi:10.1142/S0217751X21502018
[arXiv:2010.07660 [gr-qc]].

\bibitem{Noether:1918zz}
E.~Noether,
Gott. Nachr. \textbf{1918} (1918), 235-257
doi:10.1080/00411457108231446
[arXiv:physics/0503066 [physics]].

\bibitem{Komar:1958wp}
A.~Komar,
Phys. Rev. \textbf{113} (1959), 934-936
doi:10.1103/PhysRev.113.934

\bibitem{Balasin:1993fn}
H.~Balasin and H.~Nachbagauer,
Class. Quant. Grav. \textbf{10}, 2271 (1993)
doi:10.1088/0264-9381/10/11/010
[arXiv:gr-qc/9305009 [gr-qc]].
 
\bibitem{Geroch:1986jjl}
R.~P.~Geroch and J.~H.~Traschen,
Conf. Proc. C \textbf{861214} (1986), 138-141
doi:10.1103/PhysRevD.36.1017.

\bibitem{Adler:2005vn}
R.~J.~Adler, J.~D.~Bjorken, P.~Chen and J.~S.~Liu,
Am. J. Phys. \textbf{73}, 1148-1159 (2005)
doi:10.1119/1.2117187
[arXiv:gr-qc/0502040 [gr-qc]].

\bibitem{Yokoyama:2021nnw}
S.~Yokoyama,
[arXiv:2105.09676 [gr-qc]].

\bibitem{Friedmann:1924bb}
A.~Friedmann, 
Z. Phys. \textbf{21}, 326-332 (1924)
 doi :10.1007/BF01328280.
 
 \bibitem{Lemaitre:1931zza}
Georges Lemaitre, 
Mon. Not. Roy. Astron. Soc. \textbf{91}, 483-490 (1931).

\bibitem{Robertson:1935zz}
H.~P.~Robertson,
Astrophys. J. \textbf{82}, 284-301 (1935)
doi :10.1086/143681.

\bibitem{Walker:1937aa}
A.~G.~Walker,
Proceedings of the London. Mathematical Society Series 2 \textbf{42}, 90-127 (1937)
doi:10.1112/plms/s2-41.1.90.

\bibitem{Utiyama:1984bc}
R.~Utiyama,
Prog. Theor. Phys. \textbf{72} (1984), 83
doi:10.1143/PTP.72.83



\bibitem{Wald:1993nt}
R.~M.~Wald,
Phys. Rev. D \textbf{48} (1993) no.8, R3427-R3431
doi:10.1103/PhysRevD.48.R3427
[arXiv:gr-qc/9307038 [gr-qc]].

\bibitem{Townsend:1997ku}
P.~Townsend,
[arXiv:gr-qc/9707012 [gr-qc]].

\bibitem{Kodama:1979vn}
H.~Kodama,
Prog. Theor. Phys. \textbf{63} (1980), 1217
doi:10.1143/PTP.63.1217

\bibitem{Abbott:1981ff}
L.~Abbott and S.~Deser,
Nucl. Phys. B \textbf{195}, 76-96 (1982)
doi:10.1016/0550-3213(82)90049-9

 \end{thebibliography}
\end{document}